\documentclass[journal]{IEEEtran}
\usepackage[utf8]{inputenc}
\usepackage[english]{babel}
\usepackage{siunitx}
\usepackage{textcomp}
\usepackage{graphicx}
\usepackage{amsmath}
\usepackage[dvipsnames]{xcolor}
\usepackage{booktabs}
\usepackage{multirow}

\usepackage[font=small]{caption}

\ifCLASSINFOpdf

\else

\fi

\begin{document}

\title{Higher-order correlation based real-time beamforming in photoacoustic imaging}

\author{Sufayan Mulani, Souradip Paul,
         and~Mayanglambam Suheshkumar Singh % <-this % stops a space
\thanks{Sufayan Mulani, S Paul and MS Singh are with the School of Physics, Indian Institute of Science Education and Research Thiruvananthapuram, Thiruvananthapuram - 695551, Kerala, India.}

}

\maketitle

\begin{abstract}
Linear-array based photoacoustic images are reconstructed using the conventional delay-and-sum (DAS) beamforming method. Although the DAS beamformer is well suited for PA image formation, reconstructed images are often afflicted by noises, sidelobes, and other intense artifacts due to inaccurate assumptions of PA signal correlation. The work aims to develop an inversion method that reduces the occurrence of sidelobes and artifacts and improves image quality performance. We present a novel beamformer based on higher-order signal correlation, where more number of delayed PA signals are combined and summed up than the conventional delay-multiply-and-sum (DMAS). The proposed technique provides efficient improvements in resolution, contrast, and SNR compared to the traditional beamformers. Computational complexity in this method is shrunk to the same order of DAS $O(N)$. Therefore, this beamformer can be implemented in real-time PA image reconstruction. A GPU based study was performed on computation time. Proposed method almost executes in the same time frame of DAS and real-time DMAS. A validation study of the algorithm was accomplished both numerically and experimentally. Higher-order DMAS beamformers demonstrate superior reconstruction in all the cases. The quantitative evaluation of the ex-vivo phantom shows that the proposed method leads to $51\%$ and $6\%$ improvement in FWHM, $81\%$ and $39\%$ improvement in SNR compared to DAS and DMAS, respectively. Conclusively, the proposed algorithm is very much potential and promising in real-time photoacoustic imaging and its applications. 

\end{abstract}

\IEEEpeerreviewmaketitle

\section{Introduction}

High resolution and high contrast optical images have always been a great desire for research for diagnosis and monitoring various critical diseases. Over the past two decades, photoacoustic imaging (PAI) has grabbed a lot of attention from researchers due to its unique phenomenon. It is a hybrid biomedical imaging modality that renders rich optical contrast with scalable ultrasound resolution at higher imaging depth \cite{wang2012photoacoustic, zou2017biomedical}. In PAI, instantaneous absorption of transient electromagnetic (EM) radiation induces recurring thermo-elastic expansions in tissue samples. As a result, mechanical vibration is generated, and that can be detected using a single-element transducer or array transducer  \cite{suheshkumar2020fundamentals, manohar2016photoacoustics}. Light-induced PA signal reveals the distribution of various endogenous and exogenous chromophores, such as haemoglobin, melanin, nanotube, nanoparticle, e.t.c \cite{xia2014photoacoustic, steinberg2019photoacoustic, jiang2017advanced}. Therefore, functional information provided by PAI plays a promising role in diagnostic and therapeutic aspects, including cancer and other life-threatening diseases \cite{manohar2019current, attia2019review, razansky2021multi}.

Image reconstruction is the only guiding evidence for visualizing the distribution of optical absorbers, through which the entire understanding of disease or abnormality in the imaging region of interest could be opted \cite{warbal2019impact}. The objective of PA image reconstruction is to estimate the initial pressure distribution inside a tissue sample from the set of acquired PA signals. There are many reconstruction approaches proposed in a linear-array based PAI \cite{rejesh2013deconvolution, prakash2019maximum}. Delay-and-sum (DAS) is one of the most routinely used non-adaptive algorithms due to its simple implementation and real-time capability \cite{hoelen2000image}. However, it has many drawbacks such as low contrast, low resolution, wide main-lobe, higher side-lobes and other intense artifacts \cite{paul2021delay}. A vast range of studies have been performed to mitigate the limitations of the DAS beamforming algorithm \cite{pramanik2014improving}. Weighting based approaches like coherence factor (CF) \cite{wang2014snr}, modified coherence factor (MCF) \cite{mozaffarzadeh2018enhanced}, sign coherence factor (SCF) \cite{sufayan, camacho2009phase},  variational coherence factor (VCF) \cite{paul2021noise}, standard deviation based weighting factor \cite{paul2021delay, wang2018adaptive} has been utilized in PA image quality enhancement. In 2015, Matrone et al. proposed an advanced beamforming algorithm named delay-multiply-and-sum (DMAS) beamformer, which offers superior image quality compared to DAS \cite{matrone2014delay}. This improved algorithm shows excellent performance in contrast enhancement, resolution improvement and side-lobe reduction but is challenging to utilize in clinical application due to its heavy computational complexity. Mozaffarzadeh et al. suggested a modified and improved version of DMAS, called double-stage DMAS (DS-DMAS), but still computationally extensive and unable to process real-time imaging \cite{mozaffarzadeh2017double}. Later real-time DMAS was established by Jeon et al. to reduce computational overhead and extra processing time \cite{jeon2019real}. Minimum variance (MV) based adaptive beamformers, which have a wide range of applications in radar and sonar, enable to conquer the existing limitations in non-adaptive beamformers \cite{mozaffarzadeh2018linear}. A variety of modifications were investigated on MV beamformers by several research groups, such as complexity reduction, GPU optimization, shadowing suppression \cite{mozaffarzadeh2018eigenspace}. Short-lag spatial coherence (SLSC) beamformer is another option that has profound substantial promise in multiple interventional tasks in US and PA imaging like a needle-guided biopsy, backscattered echo preservation, prostate brachytherapy \cite{bell2013short}.

Highly correlated PA signals are essential to figuring out the exact target location in noisy images. Several beamforming algorithms (DMAS, DS-DMAS) have shown their potentiality to extract high signal correlation. However, the complexity of the algorithm is the primary concern in the scope of real-time imaging applications. This article addresses the higher-order correlation of DMAS beamforming, where more number delayed signals are multiplying and summing to construct strong PA signal correlations without imposing immense complexity. The algorithm is proposed with $3$, $4$ and $5$ number of delayed signals correlation named as DMAS-$3$, DMAS-$4$ and DMAS-$5$, respectively. All the multiplied signals summations present in the algorithm are expressed with a couple of simplified summations. The proposed method outperforms state-of-the-art beamformers (DAS, DMAS) in evaluating image quality metrics. Therefore, this algorithm enhances image quality and accelerates computation time towards the avenue of real-time applications. 

The remainder of this article is organized as follows. Necessary materials and methods are described in Section 2, numerical simulation and experimental results are illustrated in Section 3 and 4. Section 5 summarized the GPU implementation of the proposed method. Discussion is presented in Section 6. Finally, the conclusion of this study is given in Section 7.

\section{Materials and Methods}

\subsection{Delay and Sum (DAS) beamforming}
In DAS beamforming, the PA signals received by each transducer element are delayed according to its distance from the imaging point and then summed together to get the beamformed signal corresponding to that point. For a transducer of N element sensors, the output of the DAS beamformer is given as \cite{hoelen2000image}
\begin{equation}
    S_{DAS}=\sum_{i=1}^{N}s_i[k+\Delta_i],
    \label{das}
\end{equation}
where $s_i$ is the PA signal received by $i^{th}$ element, k is the time index and $\Delta_{i}$ is the corresponding time delay of that element. 

Although DAS beamformer has its advantages, such as simplicity and real-time processing, it suffers from high sidelobes, low resolution, low contrast and high noise.

\subsection{Delay Multiply and Sum (DMAS) beamforming}
In DMAS beamforming, similar to DAS, signals from each transducer element are delayed accordingly, but instead of directly summing these signals, they are combinatorially combined and multiplied before adding them together. Mathematically, the DMAS beamformer is given as \cite{matrone2014delay} :
\begin{equation}
    S_{DMAS}=\sum_{i=1}^{N-1}\sum_{j=i+1}^N sign(s_is_j)\times\sqrt{|s_is_j|},
    \label{DMAS}
\end{equation}
for the notation simplicity, $s_i(k+\Delta_i)$ is written as $ s_i $ and this will be followed for the rest of the article. A signed square root was taken to equate the dimensionality of output of the DMAS beamformer to RF signals without losing sign.

DMAS is a non-linear beamformer where the multiplication term acts as a correlation function, resulting in lower sidelobes, better contrast, and better resolution compared to DAS. However, the computational complexity of order $O(N^2)$, makes this beamformer challeging to achieve real-time processing. To overcome this, a simplified-DMAS beamformer was proposed, which gives the mathematically same output as DMAS but with reduced computational complexity of $O(N)$. Mathematically, it is written as:

\begin{equation}
    S_{DMAS}=\frac{\left(\sum_{i=1}^{N}sign(s_i)\sqrt{|s_i|}\right)^2 - \sum_{i=1}^{N}|s_i| }{2}.
    \label{s-DMAS}
\end{equation}

\subsection{Proposed Methods}
In this work, We consider higher order signal correlation. Instead of combinatorially combining only two terms as in DMAS, it is proposed to incorporate a higher number of terms and multiply them before the summation. The objective is to strengthen the signal and suppress the sidelobes and noises to a more significant level as multiplying terms increase. 

DMAS beamformer with $3$, $4$ and $5$ multiplying terms can be written as follows

\begin{equation}
    S_{DMAS-3}=\sum_{i=1}^{N-2}\sum_{j=i+1}^{N-1}\sum_{k=j+1}^{N} \sqrt[3]{s_is_js_k},
    \label{D3MAS}
\end{equation}

\begin{equation}
    S_{DMAS-4}=\sum_{i=1}^{N-3}\sum_{j=i+1}^{N-2}\sum_{k=j+1}^{N-1}\sum_{l=k+1}^{N} sign(s_is_js_ks_l)\times\sqrt[4]{|s_is_js_ks_l|},
    \label{D4MAS}
\end{equation}

\begin{equation}
    S_{DMAS-5}=\sum_{i=1}^{N-4}\sum_{j=i+1}^{N-3}\sum_{k=j+1}^{N-2}\sum_{l=k+1}^{N-1}\sum_{m=l+1}^{N} \sqrt[5]{s_is_js_ks_ls_m},
    \label{D5MAS}
\end{equation}
Here, powers of multiplication terms are arranged such that the output of the beamformer is correctly scaled to have the same dimensionality as the RF signal. In 4 terms DMAS, since considering only the fourth root of multiplication term would result in a loss of sign, signed fourth root is taken, similar to conventional DMAS.

Though these beamformers can produce higher-quality images, parallelly, their computational complexities are also increased to orders of $O(N^3)$, $O(N^4)$ and $O(N^5)$, making them extremely time-consuming and hard to implement. These beamformers have been simplified so that the final output of the beamformers is the same as defined above, but their computational complexity is massively reduced to an order of DAS ($O(N)$). The simplified versions of DMAS with $3$, $4$ and $5$ terms are given as

\begin{equation}
S_{DMAS-3}=\frac{1}{6}\left[ \left( \sum_{i=1}^N \sqrt[3]{s_i}\right)^3 + 2\sum_{i=1}^N s_i -3\sum_{i=1}^N \sqrt[3]{s_i}\sum_{i=1}^N (\sqrt[3]{s_i})^2\right],    
\label{S-D3MAS}
\end{equation}

\begin{equation} 
\begin{split}
S_{DMAS-4}= & \frac{1}{24}\left[ \left( \sum_{i=1}^N sign(s_i)\sqrt[4]{|s_i|}\right)^4 
\right.\\ & \left.
- 6\sum_{i=1}^N |s_i| +3\left(\sum_{i=1}^N \sqrt{|s_i|}\right)^2 \right.\\
&\left. -6\sum_{i=1}^N \sqrt{|s_i|}\left( \sum_{i=1}^N sign(s_i)\sqrt[4]{|s_i|}\right)^2 \right.\\
& \left. +8 \sum_{i=1}^N sign(s_i)\left(\sqrt[4]{|s_i|}\right)^3 \sum_{i=1}^N sign(s_i)\sqrt[4]{|s_i|}\vphantom{\left( \sum_{i=1}^N s_i\right)^4}\right], \\
\end{split}
\label{s-D4MAS}
\end{equation}

\begin{equation}
    \begin{split}
        S_{DMAS-5}= &\frac{1}{120}\left[ \left(\sum_{i=1}^N \sqrt[5]{s_i}\right)^5 +24\sum_{i=1}^N s_i  
        \right. \\ & \left. 
        -30\sum_{i=1}^N \sqrt[5]{s_i}\sum_{i=1}^N \left(\sqrt[5]{s_i}\right)^4 \right. \\
        & \left. +20\sum_{i=1}^N \left(\sqrt[5]{s_i}\right)^3\left(\sum_{i=1}^N \sqrt[5]{s_i}\right)^2 
        \right. \\& \left.
        -20\sum_{i=1}^N \left(\sqrt[5]{s_i}\right)^3\sum_{i=1}^N \left(\sqrt[5]{s_i}\right)^2 \right.\\
        & \left. +15\left(\sum_{i=1}^N \left(\sqrt[5]{s_i}\right)^2\right)^2\sum_{i=1}^N \sqrt[5]{s_i} \right.\\
        & \left. -10\sum_{i=1}^N \left(\sqrt[5]{s_i}\right)^2\left(\sum_{i=1}^N \sqrt[5]{s_i}\right)^3\right].
    \end{split}
\end{equation}

For comparison, we have shown approximate numbers of mathematical operations required for different simplified DMAS beamformers in Table \ref{tab-complex}.

\begin{table}[tbh!]
\centering
\footnotesize
\begin{tabular}{lccccc}
\hline
                        & \textbf{DAS} & \textbf{DMAS} & \textbf{DMAS-3} & \textbf{DMAS-4} & \textbf{DMAS-5}   \\ 
\hline
\textbf{Summation}      & $1N$         & $2N$          & $3N$            & $4N$            & $5N$             \\ 
\textbf{Multiplication} & $-$          & $1N$          & $2N$            & $3N$            & $4N$              \\ 
\textbf{Absolute}       & $-$          & $1N$          & $-$             & $1N$            & $-$              \\ 
\textbf{Signum}         & $-$          & $1N$          & $-$             & $1N$            & $-$              \\ 
\textbf{Square root}    & $-$          & $1N$          & $-$             & $-$             & $-$               \\ 
\textbf{Cube root}      & $-$          & $-$           & $1N$            & $-$             & $-$               \\ 
\textbf{Fourth root}    & $-$          & $-$           & $-$             & $1N$            & $-$               \\ 
\textbf{Fifth root}     & $-$          & $-$           & $-$             & $-$             & $1N$              \\ 
\hline
\end{tabular}
\caption{Approximate numbers of mathematical operations required for different beamformers.}
\label{tab-complex}
\end{table}

%%%%%%%%%%%%%%%%%%%%%%%%%%%%%%%%%%%%%%%%%%%%%%%%%%%%%%%%%%%%%%%%%%%%%%%%%%%%%%%%%%%%%%%%%%%%%%%%%%%%%%%%%%%%%%%%%%%%%%%%
\section{Numerical Simulations and Performance assessment}

\begin{figure}[tbh!]
\centering
\includegraphics[width =\columnwidth]{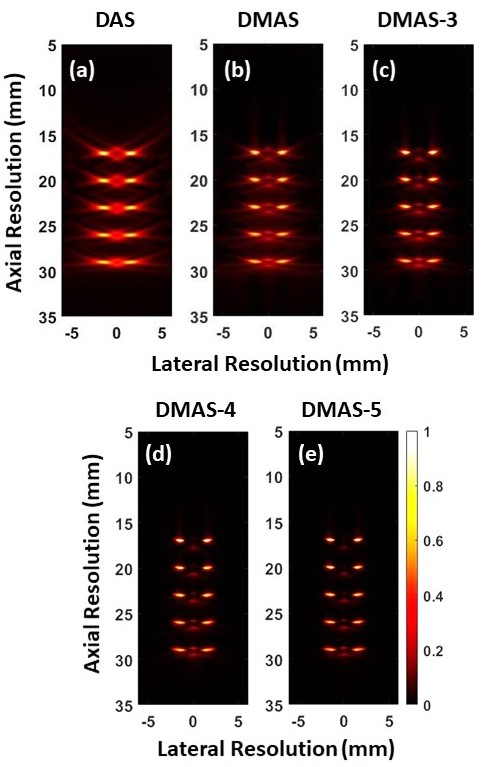}
\caption{Reconstructed PA images from detected data of $10$ simulated targets using (a) DAS, (b) DMAS (c) DMAS-3, (d) DMAS-4 and (e) DMAS-5 beamformers. }
\label{sim_res_result}
\end{figure}

\begin{figure*}[htb!]
\centering
\includegraphics[width = 17 cm]{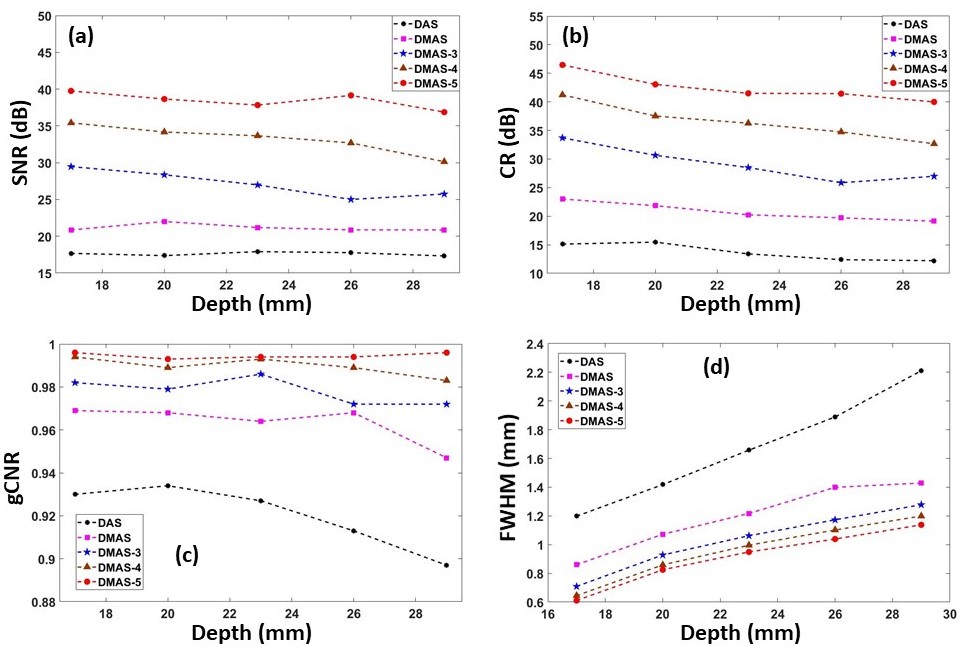}
\caption{Quantitative evaluation of DAS, DMAS and the proposed beamformers in the reconstructed images shown in Fig. \ref{sim_res_result} using (a) SNR, (b) CR (c) gCNR and (d) FWHM. }
\label{sim_res_quant_eval}
\end{figure*}

Numerical simulations were designed using the k-Wave MATLAB toolbox. For simulations, a homogeneous propagation medium of density $1500\,kg/m^3$ and sound speed $1500 \: m/s $ was considered. Five pairs of circular absorbers of radius $0.1\,mm$ were placed at the depths of $17\,mm$, $20\,mm$, $23\,mm$, $26\,mm$ and $29\,mm$ with lateral separation of $3\,mm$ between each pair. A linear array transducer of 128 elements with pitch $0.1\,mm$ was used for detecting PA signals. The central frequency of transducer and sampling frequency was taken to be $7.5\,MHz$ and $100\,MHz$, respectively. Gaussian noise with SNR of $30\,dB$ was added to each of the detected signals. Images were reconstructed using DAS, DMAS, DMAS-3, DMAS-4 and DMAS-5 beamformers and a bandpass filter was applied at the appropriate frequency range.  Envelope detection using Hilbert transformation and log compression were applied before displaying final images.

\begin{figure}[htb!]
%\centering
\includegraphics[width = 8cm]{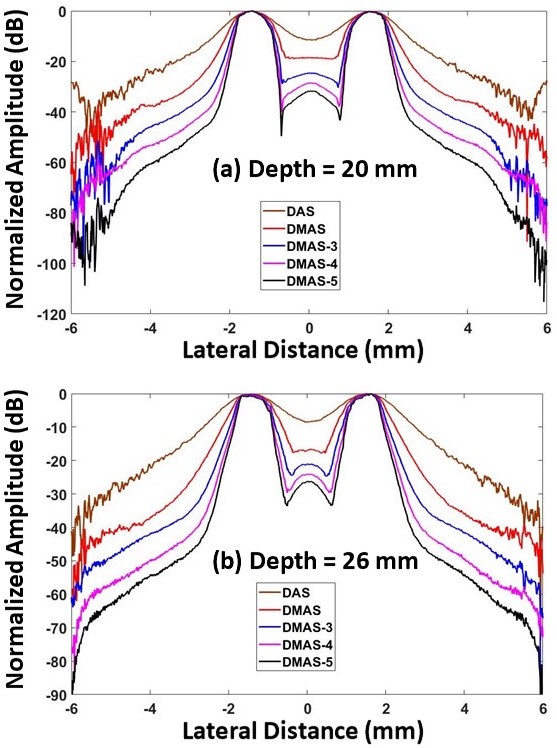}
\caption{Lateral variations of pixel values at the depths of (a) $20\,mm$ and (b) $26\,mm$ in reconstructed PA images from Fig. \ref{sim_res_result}. }
\label{sim_res_lat_var}
\end{figure}

Figure \ref{sim_res_result} shows the reconstructed images using DAS, DMAS, DMAS-3, DMAS-4 and DMAS-5. As can be seen in Fig. \ref{sim_res_result}(a), DAS beamformer provides high sidelobe, wide mainlobe and noisy images. At higher imaging depth, targets are barely visible and overlapped with sidelobes. In conventional DMAS beamformer, image quality gets slightly improved (Fig.\ref{sim_res_result}(a)). However, noise clustering and sidelobes between the targets region are still prominent. In our proposed method (Fig.\ref{sim_res_result}(c)), reconstructed images appear to be the best noise-reduced images. Further, increasing more number of signal correlations, sidelobes are suppressed, mainlobes width are decreased, and targets are better discernable (Fig.\ref{sim_res_result}(d) and (e)). To compare the beamformers in more details, lateral variations of the reconstructed images at the depths of $20\,mm$ and $26\,mm$ are presented in Fig. \ref{sim_res_lat_var}. As it is clearly demonstrated, our proposed method offers appreciable sidelobe suppression, noise reduction and narrower mainlobe compared to the conventional beamformers. Moreover, this improvement is enhanced at the increment of signals correlation. Therefore, at both the depths, $20\,mm$ and $26\,mm$, DMAS-5 beamformer has the most suppressed sidelobes and the smallest mainlobe widths, indicating the excellent performance of our proposed beamformer.

To validate the efficacy of the proposed method, we performed a quantitative comparison study of image quality metrics (CR, SNR, FWHM and gCNR). Signal-to-noise ratio (SNR) and contrast (CR) were calculated using the following formulas :
\begin{equation}
    SNR=20\log_{10}\left(\frac{\mu_{in}}{\sigma_{out}}\right),
    \label{SNR}
\end{equation}

\begin{equation}
    CR=20\log_{10}\left(\frac{\mu_{in}}{\mu_{out}}\right),
    \label{CR}
\end{equation}
 where, $\mu_{in}$ and $\mu_{out}$ are averages of PA signals over target region and background region, respectively, and $\sigma_{out}$ is the standard deviation of PA signal in the background region. SNR and CR values at the depths of $17\,mm$, $20\,mm$, $23\,mm$, $26\,mm$ and $29\,mm$ for all beamformers are presented in Figure \ref{sim_res_quant_eval} (a) and (b), where each value represents an average of two targets at that depth. As can be seen in figures both SNR and CR improve as the number of multiplying terms in DMAS are increased. DMAS-5 shows the improvement of around $25\, dB$ and and $20\, dB$ at all depths in SNR compared to DAS and DMAS, respectively. We can see improvement of CR in DMAS-5 around $30\,dB$ compared to DAS and $25\,dB$ compared to DMAS. Full-width-half-maxima (FWHM) at different depths are presented in Figure \ref{sim_res_quant_eval} (d). It can be seen that, at all the depths, FWHM value decreases as we go from DAS to DMAS-5. A newly proposed image quality metric called generalized contrast-to-noise ratio (gCNR) was also calculated for all  reconstructed images.
 
\begin{equation}
    gCNR=1-\sum_{x=0}^{1}min\left\{ h_i (x), \, h_o(x)\right\},
    \label{gcnr}
\end{equation}
where, x is the normalised signal amplitude, $h_i$ and $h_o$ are the probability density functions of signal amplitudes within target and background regions, respectively. Calculated gCNR are presented in Figure \ref{sim_res_quant_eval} (c). Our proposed methods show improvement in gCNR. DMAS-5 results in images with gCNR of $0.996$, $0.993$, $0.994$, $0.994$ and $0.996$ at depths of $17\,mm$, $20\,mm$, $23\,mm$, $26\,mm$ and $29\,mm$, respectively.  

\begin{figure}[htb!]
\centering
\includegraphics[width = \columnwidth]{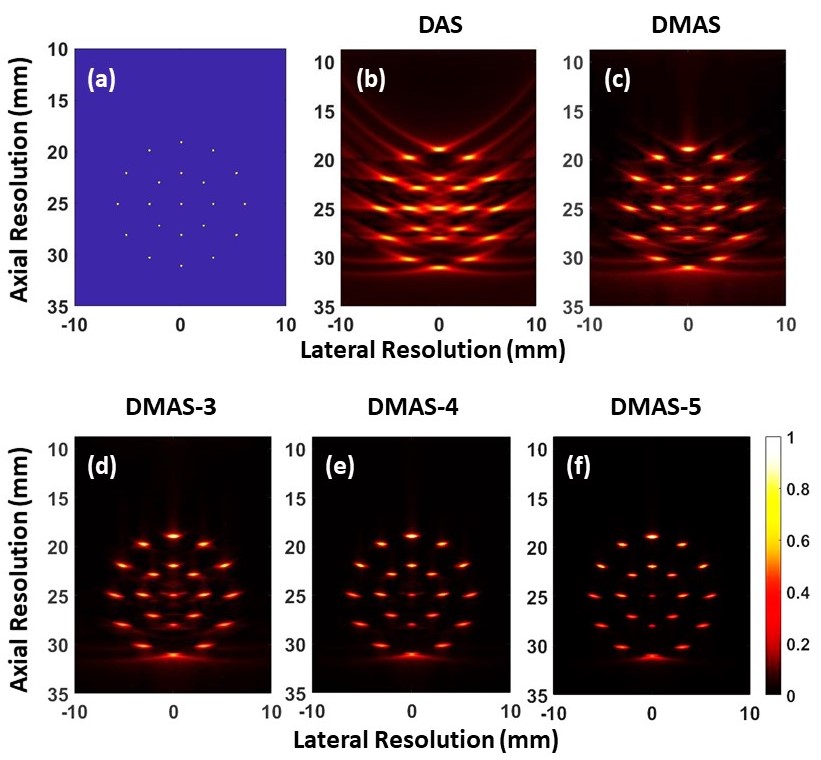}
\caption{(a) Ground truth image of the complex simulated phantom, and the Reconstructed PA images from detected data of simulated phantom using (b) DAS, (c) DMAS (d) DMAS-3, (e) DMAS-4 and (f) DMAS-5 beamformers. }
\label{sim_comp_result}
\end{figure}

\begin{figure}[htb!]
\centering
\includegraphics[width = \columnwidth]{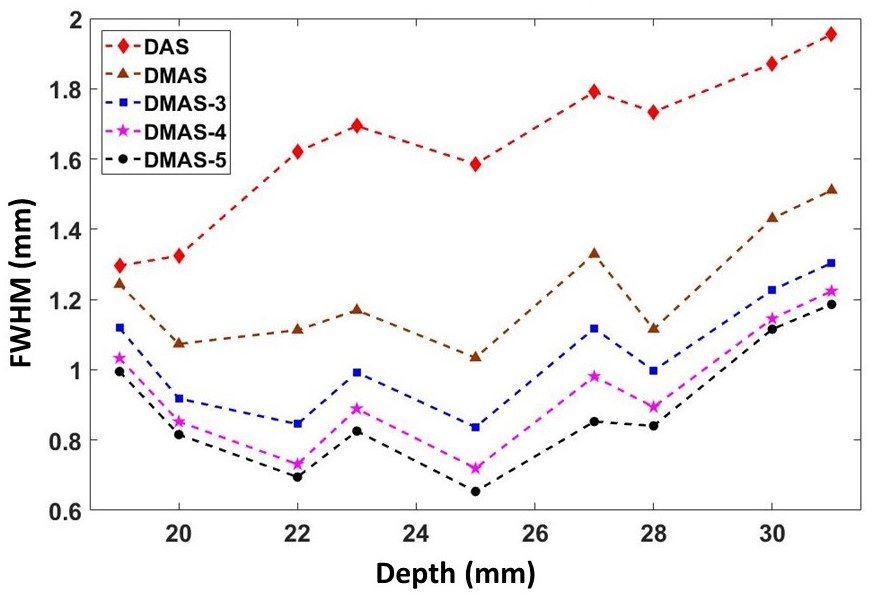}
\caption{FWHM values at different depths in reconstructed images of complex simulation phantom (Fig. \ref{sim_comp_result}). }
\label{sim_comp_fwhm}
\end{figure}

\begin{figure*}[htb!]
\centering
\includegraphics[width = 14 cm]{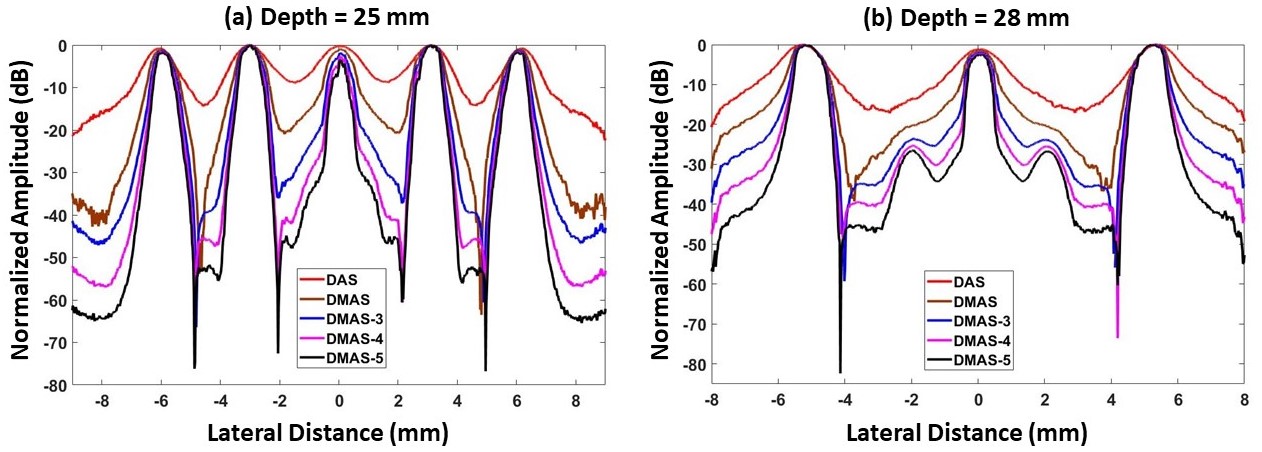}
\caption{Lateral variations of pixel values at the depths of (a) $25\,mm$ and (b) $28\,mm$ in reconstructed PA images from Fig. \ref{sim_comp_result}. }
\label{sim_comp_lat_var}
\end{figure*}

The proposed beamformers were evaluated on a more complex simulation. Twenty one circular absorbers of radius $0.1\,mm$ were placed on two concentric circles of radii $30\,mm$ and $60\,mm$. Eight targets were placed on inner circle with angular difference of $45$ degrees, twelve targets were placed on outer circle with angular difference of $30$ degrees and one target was placed at the center. All other parameters were kept same as before. Fig. \ref{sim_comp_result} exhibits the reconstructed images. It can be seen that DAS method is highly corrupted with strong sidelobes and noises (Fig. \ref{sim_comp_result}(a)). DMAS beamformer improves image quality in terms of sidelobe suppression, however smearing phenomenon is still noticeable in Fig. \ref{sim_comp_result}(b). From Fig. \ref{sim_comp_result}(c)-(e), it is obvious that higher order DMAS outperforms the conventional DAS and DMAS beamformers. Targets are nicely detectable and remarkably improved owing to the reduction of sidelobes around the targets. Lateral profile of the beamformed responses at the depths of $25\,mm$ and $28\,mm$ are shown in Fig. \ref{sim_comp_lat_var} (a) and (b), respectively. It is comprehensible that proposed methods suppress sidelobes more efficiently than conventional DAS and DMAS beamformers. At the depth of $25\,mm$, DMAS-5 suppress sidelobes by around $40\,dB$ compared to DAS and around $20\,dB$ compared to DMAS. DMAS-5 also results in the sharpest peaks compared to all other beamformers, demonstrating that the proposed method provides better resolution. FWHM values at different depths are presented in Figure \ref{sim_comp_fwhm}, where each value is the average of FWHM values of all targets at that depth. For example, at the depth of $25\,mm$ five targets are present, so the value presented in figure at the depth of $25\,mm$ is average of these five targets, whereas at the depth of $19\,mm$ only one target is present, hence values corresponding to the depth of $19\,mm$ are the FWHM of that target. DMAS-5 results in lowest FWHM values at all the depths compared to all other beamformers.

\section{Experimental Study}
\begin{figure}[htb!]
\centering
\includegraphics[width = \columnwidth]{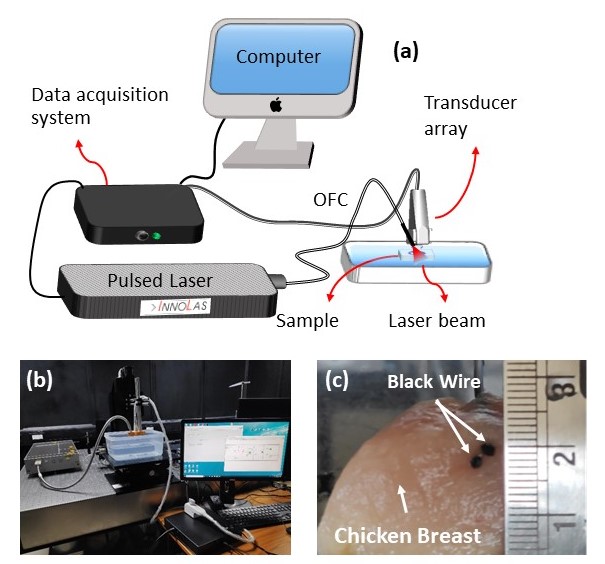}
\caption{(a) schematic diagram of our home-built linear array based photoacoustic tomogarphy (PAT) set-up and corresponding photograph (b); (c) imaging sample of $2$ black wire targets embedded in chicken breast }
\label{exp_setup}
\end{figure}

An experimental study was conducted to further validate the efficacy of proposed beamformers. A home built PAT system (see Fig. \ref{exp_setup} (a) and (b)) with $16$ elements linear array transducer operating at $3\,MHz$ central frequency and  $1.2\, mm$ pitch width was used to acquire the PA signal. An ex-vivo phantom was designed by placing two black wires with a circular cross-section of diameter $1.65\,mm$, inside a chicken breast tissue of size $30\,mm\times 30\,mm \times 15\,mm$ (Fig. \ref{exp_setup} (c)). Lateral separation between the two wires was kept at $3\,mm$, and axial separation at $2.5\,mm$. 

\begin{figure}[htb!]
\centering
\includegraphics[width = \columnwidth]{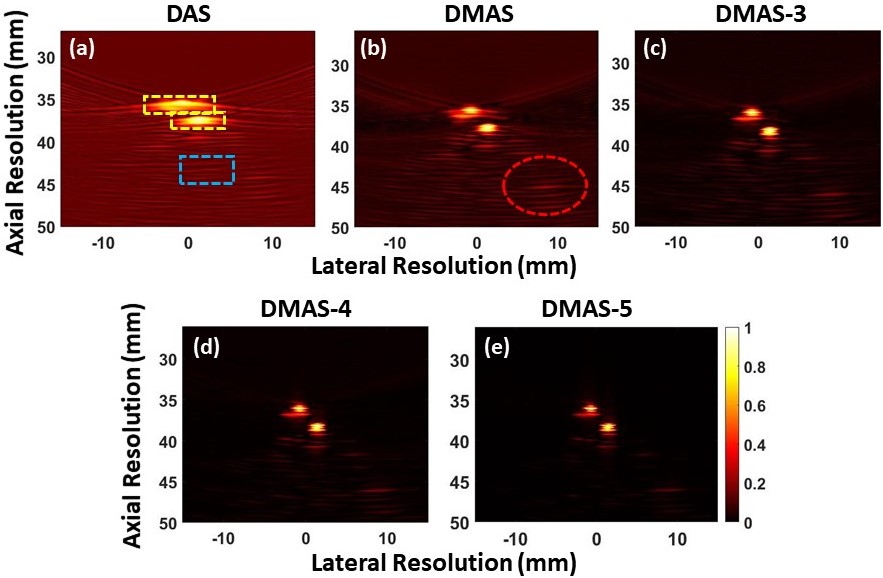}
\caption{Reconstructed PA images from the experimental data using (a) DAS, (b) DMAS (c) DMAS-3, (d) DMAS-4 and (e) DMAS-5 beamformers. }
\label{exp_result}
\end{figure}

Reconstructed images are presented in Fig. \ref{sim_comp_result}. As can be seen, the image reconstructed using DAS (Fig. \ref{sim_comp_result} (b)) is noisy and has high sidelobes and wide mainlobe width. DMAS suppresses sidelobes to an extent, but they are still visible, as are many other artifacts. DMAS-3 and DMAS-4 further improve the image quality by increasing contrast and resolution (see Fig. \ref{sim_comp_result} (d) and (e)). Out of all the beamformers, DMAS-5 gives the best results. In the image reconstructed using DMAS-5 (Fig. \ref{sim_comp_result} (f)), noises and sidelobes have almost disappeared, some artifacts are still present, but they have diminished, contrast and resolution are also improved. For example, in the image reconstructed using DMAS (see Fig. \ref{exp_result} (b)), the region marked with a red circle has different artifacts that diminish as higher-order DMAS beamformers are used. It is evident that the proposed method gives the best results compared to other conventional methods.

\begin{figure*}[htb!]
\centering
\includegraphics[width = 14 cm]{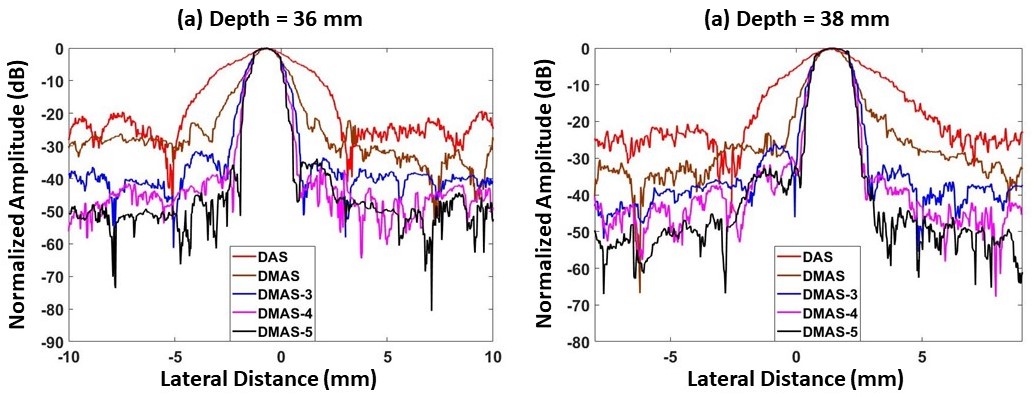}
\caption{Lateral variations of pixel values at the depths of (a) $36\,mm$ and (b) $38\,mm$ in reconstructed PA images from Fig. \ref{exp_result}}
\label{exp_lat_var}
\end{figure*}

Lateral variations in pixel values at the depths of 36mm and 38mm for different beamformers are presented in Fig. \ref{exp_lat_var}. The proposed beamformers have significantly improved lateral profiles compared to DAS and DMAS. In the lateral profile of the proposed beamformers, sidelobe suppression is very noticeable. Mianlobe width also decreases as the number of multiplying terms in the DMAS beamformer increases.

\begin{figure}[htb!]
\centering
\includegraphics[width = \columnwidth]{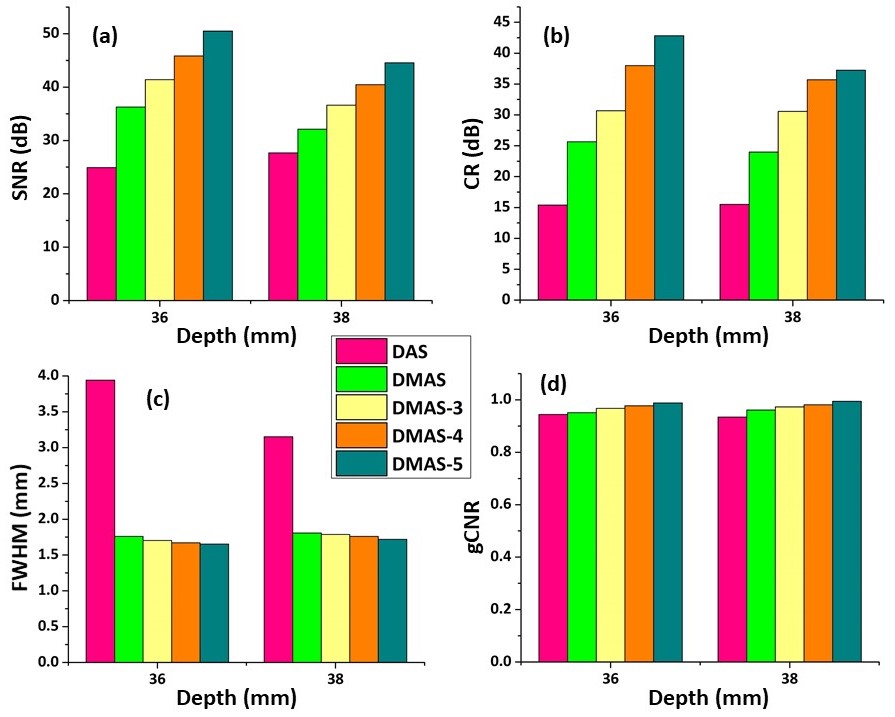}
\caption{Quantitative evaluation of DAS, DMAS and the proposed beamformers in the reconstructed images shown in Fig. \ref{exp_result} using (a) SNR, (b) CR (c) FWHM and (d) gCNR. }
\label{exp_quant_eval}
\end{figure}

SNR, CR, gCNR and FWHM were calculated for different beamformers in a similar way as done in the simulation study and presented in Fig. \ref{exp_quant_eval}. Target and background regions are shown in yellow and blue rectangles, respectively. Proposed beamformers show improvements in all the metrics. DMAS-5 leads to an increase in SNR of around 25 dB at a depth of 36mm and 17dB at a depth of 38mm, compared to DAS. CR and gCNR values also show that the proposed beamformers outperform DAS and DMAS. An interesting thing to note is that DMAS-5 leads to an FWHM value of 1.6mm at a depth of 36mm and 1.7mm at a depth of 38mm, which is very close to the actual diameter of the wire target.

\section{GPU implementation}
Real-time processing is crucial in various clinical and pre-clinical applications. Even though the complexity of the proposed beamformers has been reduced to an order of $O(N)$, processing time is still higher than DAS due to the additional mathematical operations required. Therefore GPU based parallel computing was performed for achieving better temporal resolution. GPUs are made up of thousands of tiny cores specifically designed to do multiple tasks simultaneously. As a result, highly parallel algorithms, such as the proposed beamformers, take less time for processing on the GPU than on the CPU. For processing, NVIDIA GeForce GT 730 GPU with $384$ CUDA cores was used along with Intel Core i7-7700 CPU. The programming was done in compute unified device architecture (CUDA) using Microsoft Visual Studio. In the programme, before calling the CUDA kernel, all the delays ($\Delta_i$) corresponding to each pixel and transducers were calculated and stored in the GPU (global) memory to reduce the processing time. Table \ref{tab:GPU time} presents the average processing time based on $50$ runs of DAS, DMAS and simplified versions of proposed methods on a GPU. The table shows that the DMAS-5 would generate around $12$ frames per second of size $2048\times2048$. As a result, our proposed method has been shown to be completely suited for real-time imaging applications. 

\begin{table}[htb!]
\centering
\small
\begin{tabular}{cccccc}
\hline
\multirow{2}{*}{\begin{tabular}[c]{@{}l@{}}\textbf{Image Size}\\ \textbf{(pixel$\times$pixel)}\end{tabular}} & \multicolumn{5}{l}{\textbf{Reconstruction Time (ms)}} \\
\cline{2-6}
                        & \textbf{DAS}    & \textbf{DMAS}   & \textbf{DMAS-3} & \textbf{DMAS-4} & \textbf{DMAS-5} \\
\hline
512$\times$512 & 2.912  & 3.028  & 3.309  & 4.4    & 5.133  \\
1024$\times$1024      & 11.051 & 11.6   & 12.718 & 17.121 & 20.03  \\
2048$\times$2048      & 43.438 & 45.784 & 50.205 & 68.006 & 79.653 \\
\hline
\end{tabular}
\caption{Average processing time required for each beamforming method in a GPU}
\label{tab:GPU time}
\end{table}

\section{discussion}
We introduced a higher-order correlation-based beamformer to enhance the performance of PA imaging. The main advantages of this method are resolution improvement, clustering noise and sidelobe suppression compared to conventional methods like DAS and DMAS. Simulation and experimental results are proper evidence of this image quality enhancement. DAS beamformer sustains various drawbacks in the reconstructed images due to its blindness and non-adaptiveness. Though some newly introduced beamforming methods, such as double-stage DMAS and minimum variance-based beamformers, produce better results than conventional beamformers, these beamformers have significantly higher computational complexity. As a result, implementing these beamformers in real-time imaging is extremely difficult, and they are consequently unsuitable for clinical usage. To make the proposed beamformers real-time applicable, they have been simplified; their complexity is reduced to the same order of DAS. The reconstructed images (Fig. \ref{sim_res_result}, \ref{sim_comp_result} and \ref{exp_result}) exhibit that our proposed methods outperform existing methods. In Fig. \ref{sim_res_result}, targets are better distinguishable and detectable at higher imaging depth using our technique. As signal correlation increases, clustering noises around the targets are highly suppressed. Lateral variations in Fig. \ref{sim_res_lat_var}, \ref{sim_comp_lat_var} and \ref{exp_lat_var} show that sidelobes and the widths of the mainlobes have been significantly lowered in higher-order DMAS. Quantitative assessments of image quality are presented in Fig. \ref{sim_res_quant_eval} and \ref{exp_quant_eval}, which provides superiority of the algorithm in image quality characterization. 

The unique potentiality of the proposed algorithm put forward its execution time in the same benchmark of DAS and real-time DMAS. It also enables to enhance image quality with low accessible hardware. Though this method takes extra processing time due to the additional memory access for absolute, signum, square and Cube root operation, the slowness of processing time will not influence the real-time implementation. The reduced computational burden will facilitate hastening performance that could bring down the price of PAI systems by reducing the requirements of ultrasound detectors and associated hardware, including the DAQ system. Several clinical trials have been performed in linear array-based PA imaging, and they acknowledge its capability in cancer imaging, chronic inflammatory diseases and lymph node mapping. Since the main focus of the proposed method is also performance improvement of similar low-cost linear array probes, we believe that this contribution will be beneficial for better clinical outgrowth. Recently, the utilization of Light-emitting diode (LED) based PA imaging systems has been massively expanding due to its affordability and versatility. Therefore, this method can be smoothly integrated into LED-based PA systems. Owing to the noise reduction capability of the algorithm, it will be fruitful to overcome the limitation of the LED-based PAI, which are inherently noisy and provides lower SNR than traditional optical resonator-based laser systems. Further, This method can be combined with coherence based real-time weighting factors to improve the final results. In this paper, we have simplified proposed beamformers only to the order of $5$, but this can be done for higher orders as well. However, since these beamformers are so sensitive to coherent signals, increasing the order of DMAS beamformers (greater than 5) may result in signal loss. Further study needs to be done to learn the performance of higher-order DMAS beamformers. Overall, this method certifies a new insight towards the advancement of linear array PAI and better clinical outputs.

\section{Conclusion}

In this paper, higher-order optimized DMAS beamformers are presented to overcome the persisting drawbacks in state-of-the-art methods. The proposed algorithms were applied and evaluated in both simulated and experimental data. Both simulation and experimental results confirmed the potentiality of these methods in image quality improvement. Additionally, these beamformers do not add significantly to the computational complexity as compared to traditional methods. Processing time has been verified with partially parallelized GPU based implementation. Results show that the time scales for execution are comparable to DAS and DMAS. Therefore, these powerful algorithms exhibit satisfactory robustness in real-time PAI and could be utilized for the betterment of clinical outcomes.

\bibliographystyle{ieeetr}
\bibliography{main.bib}

\end{document}